\documentclass{article}
\usepackage{cite}
\usepackage{graphicx}
\usepackage{dcolumn}

\begin{document}

\title{On the Hellmann-Feynman theorem for degenerate states}
\author{Francisco M. Fern\'{a}ndez\thanks{%
fernande@quimica.unlp.edu.ar} \\
INIFTA, DQT, Sucursal 4, C.C 16, \\
1900 La Plata, Argentina}
\maketitle

\begin{abstract}
In this paper we discuss the validity of the Hellmann-Feynman theorem (HFT)
for degenerate states. We derive it in a general way and apply it to simple
illustrative examples. We also analyze a recent paper that shows results
that apparently suggest that the HFT does not apply to degenerate states.
\end{abstract}

\section{Introduction}

\label{sec:intro}

Many years ago Feynman\cite{F39} developed a method for the calculation of
forces in molecules that does not require the explicit use of the derivative
of the energy. This expression, known as the Hellmann-Feynman theorem (HFT),
is discussed in almost every book on quantum mechanics and quantum chemistry%
\cite{CDL77,P68} and some pedagogical articles discuss its utility in
quantum mechanics\cite{E54,BS89}. It is worth mentioning its application to
perturbation theory\cite{E54}, even for degenerate states\cite{BS89}.

Some time ago Zhang and George\cite{ZG02} reported a supposedly failure of
the theorem in the case of degenerate states and proposed a remedy. Such
assessment resulted curious in the light that the proof of the theorem does
not require that the states are nondegenerate\cite{F39,CDL77,P68,E54,BS89}.
Several authors commented on this paper proving Zhang and George wrong with
respect to the failure of the HFT\cite{AC03,F04,V04,BHM04}. In particular,
Fern\'{a}ndez\cite{F04} showed that the expression for the supposed remedy
is correct but unnecessary because the original diagonal HFT is valid for
degenerate states provided that one chooses the correct linear combinations
of the degenerate eigenfunctions for the calculation.

In a recent paper Roy and Sharma\cite{RS19} argue that the HFT is not valid
for degenerate states and, curiously, look for support from those articles
that draw the opposite conclusion\cite{AC03,F04,V04,BHM04}. In particular,
these authors stress the fact that the HFT exhibits discontinuities at the
crossings between energy levels. It is worth mentioning that Alon and
Cederbaum\cite{AC03} and Fern\'{a}ndez\cite{F04} already pointed out that
there are no such discontinuities but their conclusions seem to have been
misinterpreted.

In the light of the results derived by Roy and Sharma\cite{RS19} it seems
necessary to discuss the HFT for degenerate states with more detail. In Sec.~%
\ref{sec:HFT} we derive the HFT and discuss its validity for degenerate
states. We also show that the application of group theory enables one to
completely bypass the problem posed by degenerate states. In Sec.~\ref
{sec:Examples} we illustrate the main theoretical conclusions by means of
simple models. Finally, in Sec.~\ref{sec:Conclusions} we summarize the main
results and draw conclusions.

\section{The Hellmann-Feynman theorem}

\label{sec:HFT}

In this section we do not merely follow the main arguments given in our
earlier paper\cite{F04} but provide much more information that we deem
useful for a better understanding of the problem. If the Hamiltonian
operator $H(\lambda )$ depends on a parameter $\lambda $ then its
eigenvalues $E_{n}$ and eigenfunctions $\psi _{n}$ will also depend on this
parameter. For simplicity we assume that $\left\langle \psi _{m}\right|
\left. \psi _{n}\right\rangle =\delta _{mn}$ for all $\lambda $. If we
differentiate $H\psi _{n}=E_{n}\psi _{n}$ with respect to $\lambda $ and
apply $\left\langle \psi _{m}\right| $ from the left we obtain
\begin{equation}
\left\langle \psi _{m}\right| \frac{\partial H}{\partial \lambda }\left|
\psi _{n}\right\rangle +\left\langle \psi _{m}\right| H\left| \frac{\partial
\psi _{n}}{\partial \lambda }\right\rangle =\frac{\partial E_{n}}{\partial
\lambda }\delta _{mn}+E_{n}\left\langle \psi _{m}\right| \left. \frac{%
\partial \psi _{n}}{\partial \lambda }\right\rangle .
\label{eq:derivative_schro}
\end{equation}
If we take into account that
\begin{equation}
\left\langle \psi _{m}\right| H\left| \frac{\partial \psi _{n}}{\partial
\lambda }\right\rangle =\left\langle H\psi _{m}\right. \left| \frac{\partial
\psi _{n}}{\partial \lambda }\right\rangle =E_{m}\left\langle \psi
_{m}\right. \left| \frac{\partial \psi _{n}}{\partial \lambda }\right\rangle
,
\end{equation}
then equation (\ref{eq:derivative_schro}) becomes a general expression for
the HFT
\begin{equation}
\left\langle \psi _{m}\right| \frac{\partial H}{\partial \lambda }\left|
\psi _{n}\right\rangle =\frac{\partial E_{n}}{\partial \lambda }\delta
_{mn}+\left( E_{n}-E_{m}\right) \left\langle \psi _{m}\right| \left. \frac{%
\partial \psi _{n}}{\partial \lambda }\right\rangle .  \label{eq:HFT_gen}
\end{equation}
When $m=n$ we obtain the well known diagonal form of the HFT\cite{F39}
\begin{equation}
\left\langle \psi _{n}\right| \frac{\partial H}{\partial \lambda }\left|
\psi _{n}\right\rangle =\frac{\partial E_{n}}{\partial \lambda }.
\label{eq:HFT_diag}
\end{equation}
Notice that the proof of the HFT does not assume that the states are
nondegenerate; in fact in the degenerate case equation (\ref{eq:HFT_gen})
becomes
\begin{equation}
\left\langle \psi _{m}\right| \frac{\partial H}{\partial \lambda }\left|
\psi _{n}\right\rangle =0,\;E_{m}=E_{n},\;m\neq n.  \label{eq:H'_mn=0}
\end{equation}
Obviously, in the case of degenerate states we have to take into account
both equations (\ref{eq:HFT_diag}) and (\ref{eq:H'_mn=0}) simultaneously.

Suppose that at $\lambda =\lambda _{0}$ the energy level $E_{n}$ is $g_{n}$%
-fold degenerate
\begin{equation}
H\varphi _{n+i}=E_{n}\varphi _{n+i},\;\left\langle \varphi _{n+i}\right|
\left. \varphi _{n+j}\right\rangle =\delta _{ij},\;i,j=0,1,\ldots ,g_{n}-1.
\label{eq:degenerate_level}
\end{equation}
Any linear combination of the eigenfunctions $\varphi _{n+i}$
\begin{equation}
\psi _{n+i}=\sum_{j=0}^{g_{n}-1}c_{ji}\varphi _{n+j},
\end{equation}
will be eigenfunction of $H$ with eigenvalue $E_{n}$. However, a set of $g_n$
linearly independent linear combinations will not necessarily satisfy the
HFT unless the coefficients $c_{ji}$ are chosen so that
\begin{equation}
\left\langle \psi _{n+i}\right| \frac{\partial H}{\partial \lambda }\left|
\psi _{n+j}\right\rangle =\frac{\partial E_{n+i}}{\partial \lambda }\delta
_{ij},\;i,j=0,1,\ldots ,g_{n}-1,  \label{eq:H'_ij}
\end{equation}
in agreement with equations (\ref{eq:HFT_diag}) and (\ref{eq:H'_mn=0}).
Notice, for example that the arbitrary eigenfunctions
\begin{equation}
\varphi _{n+i}=\sum_{j=0}^{g_{n}-1}c_{ij}^{*}\psi _{n+j},
\end{equation}
will not satisfy the diagonal HFT
\begin{equation}
\left\langle \varphi _{n+i}\right| \frac{\partial H}{\partial \lambda }%
\left| \varphi _{n+i}\right\rangle =\sum_{j=0}^{g_{n}-1}\left| c_{ij}\right|
^{2}\frac{\partial E_{n+j}}{\partial \lambda },  \label{eq:H'_ii_varphi}
\end{equation}
unless all the slopes are equal: $\partial E_{n+j}/\partial \lambda
=\partial E_{n}/\partial \lambda $. This situation occurs, for example, when
the variation of $\lambda $ does not change the symmetry of the problem and
the degeneracy is not removed. It is clear that the diagonal elements of $%
dH/d\lambda $ calculated with arbitrary degenerate eigenfunctions of $H$ at $%
\lambda =\lambda _{0}$ will simply yield averages of the actual slopes of
the eigenvalues. The actual slopes are given by those eigenfunctions that
satisfy equation (\ref{eq:H'_ij}). It is obvious that this condition can
always be satisfied because the coefficients $c_{ji}$ are given by a
straightforward diagonalization of the $g_{n}\times g_{n}$ matrix
representation of $dH/d\lambda $ at $\lambda =\lambda _{0}$.

Although these arguments were clearly stated in our earlier paper\cite{F04},
Roy and Sharma\cite{RS19} recently suggested that the HFT breaks down at
degeneracies in the energy spectrum and that this fact explains the
discontinuities of $I_{c}$ and $I_{c}^{var}$ that they obtained under such
conditions. However, it has clearly been shown that not only is the HFT
strictly valid at degeneracies but that there is no discontinuity whatsoever%
\cite{AC03,F04}. In fact, the degenerate eigenfunctions that satisfy
equations (\ref{eq:HFT_diag}) and (\ref{eq:H'_mn=0}) at $\lambda _{0}$ are
given by the continuity equation
\begin{equation}
\psi _{n}\left( \lambda _{0}\right) =\lim\limits_{\lambda \rightarrow
\lambda _{0}}\psi _{n}(\lambda ),  \label{eq:continuity}
\end{equation}
and all the mathematical relationships, like (\ref{eq:HFT_gen}) for example,
are continuous at $\lambda _{0}$. It can be shown that the discontinuities
in $I_{c}$ and $I_{c}^{var}$ found by Roy and Sharma\cite{RS19} have a
completely different origin and that any discrepancy between the left and
right hand sides of equation (\ref{eq:HFT_diag}) is the result of a wrong
choice of the eigenfunctions at the level crossings.

The results above apply to exact eigenfunctions and the question arises
about their validity in approximate calculations. In order to illustrate
this point we assume that we resort, for example, to the Rayleigh-Ritz
variational method with an orthonormal basis set $\left\{ f_{1},f_{2},\ldots
,f_{N}\right\} $. In this case we look for approximate eigenfunctions
\begin{equation}
\eta _{n}=\sum_{j=1}^{N}c_{jn}f_{j},  \label{eq:RR_lin_comb}
\end{equation}
that lead to a diagonal matrix representation of the Hamiltonian
\begin{equation}
\left\langle \eta _{m}\right| H\left| \eta _{n}\right\rangle =W_{n}\delta
_{mn},\;m,n=1,2,\ldots ,N,  \label{eq:RR_H_mn}
\end{equation}
where $W_{n}$ is expected to be an approximation to $E_{n}$. In the case of
degenerate solutions
\begin{equation}
W_{n+i}=W_{n},\;i=0,1,\ldots ,g_{n}-1,  \label{eq:RR_degeneracy}
\end{equation}
we choose the linear combinations that also satisfy
\begin{equation}
\left\langle \eta _{m}\right| \frac{dH}{d\lambda }\left| \eta
_{n}\right\rangle =\frac{dW_{n}}{d\lambda }\delta _{mn},\;m,n=1,2,\ldots
,g_{n}-1.  \label{eq:RR_H'_mn}
\end{equation}
One can easily convince oneself that it is always possible to obtain linear
combinations (\ref{eq:RR_lin_comb}) that satisfy both conditions (\ref
{eq:RR_H_mn}) and (\ref{eq:RR_H'_mn}).

In many cases one can avoid all the problems just mentioned by simply
resorting to group theory\cite{C90} and choosing symmetry-adapted basis sets%
\cite{AF19}. If the basis functions $f_{j}$ are adapted to the symmetry of
the problem we can treat each irreducible representation (irrep) as an
independent problem. Since states of the same symmetry do not cross\cite{F14}
(and references therein) the prescription for the choice of suitable
degenerate eigenfunctions mentioned above is bypassed. If $E_{m}(\lambda )$
and $E_{n}(\lambda )$ cross at $\lambda =\lambda _{0}$ then the
corresponding states $\psi _{m}$ and $\psi _{n}$ have different symmetry and
automatically satisfy equation (\ref{eq:H'_mn=0}). If the dimension of a
given irrep is greater than one the energies of the degenerate states have
the same slope and any linear combination of those states satisfies the HFT.
It is worth mentioning that if a group of unitary operators $\left\{
U_{0},U_{1},\ldots ,U_{K}\right\} $ leave the Hamiltonian invariant $%
U_{i}^{\dagger }HU_{i}=H$ for all $\lambda $ then they also leave $%
dH/d\lambda $ invariant.

The results of this section should be applied carefully to the calculation
of the persistent current carried out by Roy and Sharma\cite{RS19}. They
consider a one-particle Hamiltonian $H$ with eigenvalues $\epsilon _{n}(\phi
)$, $-N/2\leq n<N/2$. The lowest energy of their independent-fermions model
is given by an expression of the form\cite{RS19}
\begin{equation}
E_{0}(\phi )=\sum_{n}\epsilon _{n}(\phi )\theta \left( E_{F}-\epsilon
_{n}\right) ,  \label{eq:RS_E_0}
\end{equation}
where $E_{F}$ is the Fermi energy and $\theta (x)$ the Heaviside step
function. This expression is unclear and its straightforward application may
lead to a discontinusous function $E_{0}(\phi )$. In practice, the authors
apparently consider a fixed number $N_{p}$ of fermions and show results for
different values of $\nu =N_{p}/N$. Consequently, $E_{0}(\phi )$ is
continuous but will have a discontinuous derivative $dE_{0}/d\phi $ at
crossing points. In such cases the HFT still applies to each piece of a
piecewise-defined function and most care should be taken at the joints.

We will illustrate these theoretical results by means of simple examples in
Sec.~\ref{sec:Examples}.

\section{Examples}

\label{sec:Examples}

Our first example is the one discussed in our earlier paper\cite{F04}:

\begin{equation}
\hat{H}=\frac{1}{2}(\hat{p}_{x}^{2}+\hat{p}_{y}^{2})+\frac{\omega ^{2}}{2}(%
\hat{x}^{2}+\hat{y}^{2})+\lambda \hat{x}\hat{y},  \label{eq:HO}
\end{equation}
where $[\hat{x},\hat{p}_{x}]=[\hat{y},\hat{p}_{y}]=i$ and all other
commutators between coordinates and momenta being zero. The Schr\"{o}dinger
equation is separable in terms of the coordinates
\begin{equation}
q_{1}=\frac{1}{\sqrt{2}}\left( x+y\right) ,\;q_{2}=\frac{1}{\sqrt{2}}\left(
x-y\right) .  \label{eq:q1,q2}
\end{equation}
The eigenvalues and eigenfunctions are given by
\begin{eqnarray}
E_{mn} &=&\left( m+\frac{1}{2}\right) \sqrt{k_{1}}+\left( n+\frac{1}{2}%
\right) \sqrt{k_{2}},\;m,n=0,1,\ldots ,  \nonumber \\
k_{1} &=&\omega ^{2}+\lambda ,\;k_{2}=\omega ^{2}-\lambda ,  \nonumber \\
\psi _{mn} &=&\phi _{m}\left( k_{1},q_{1}\right) \phi _{n}\left(
k_{2},q_{2}\right) ,  \label{eq:HO_eigens}
\end{eqnarray}
where $\phi _{m}\left( k,q\right) $ is an eigenfunction of the harmonic
oscillator $H_{HO}=p_{q}^{2}/2+kq^{2}/2$.

The eigenvalues and eigenfunctions in equation (\ref{eq:HO_eigens}) satisfy
the diagonal HFT for all $\lambda $
\begin{equation}
\frac{\partial E_{mn}}{\partial \lambda }=\left\langle \psi _{mn}\right|
xy\left| \psi _{mn}\right\rangle =\frac{2m+1}{4\sqrt{k_{1}}}-\frac{2n+1}{4%
\sqrt{k_{2}}}.  \label{eq:dEmn/lam(lam)}
\end{equation}
When $\lambda =\lambda _{0}=0$ the energy levels with $m+n=\nu $ are $(\nu
+1)$--fold degenerate and it follows from equation (\ref{eq:dEmn/lam(lam)})
that
\begin{equation}
\left. \frac{\partial E_{mn}}{\partial \lambda }\right| _{\lambda
=0}=\left\langle \psi _{mn}\right| xy\left| \psi _{mn}\right\rangle
_{\lambda =0}=\frac{m-n}{2\omega }.  \label{eq:dEmn/dlam(lam0)}
\end{equation}

The alternative degenerate eigenfunctions of $H(\lambda =0)$%
\begin{equation}
\varphi _{\nu i}(x,y)=\phi _{\nu -i}(\omega ,x)\phi _{i}(\omega
,y),\;i=0,1,\ldots ,\nu ,  \label{eq:HO_varphi}
\end{equation}
do not satisfy the HFT at $\lambda =0$ except for $m=m$ because
\begin{equation}
\left\langle \varphi _{\nu i}\right| xy\left| \varphi _{\nu i}\right\rangle
=0,\;i=0,1,\ldots ,\nu .
\end{equation}
The eigenfunctions that satisfy the HFT at $\lambda =0$ are given by the
continuity equation (\ref{eq:continuity})
\begin{equation}
\psi _{mn}(\lambda =0)=\lim\limits_{\lambda \rightarrow 0}\psi _{mn}(\lambda
)=\phi _{m}\left( \omega ,q_{1}\right) \phi _{n}\left( \omega ,q_{2}\right) .
\label{eq:HO_psi_mn(0)}
\end{equation}
We appreciate that there is neither ambiguity nor discontinuity in the HFT
in the case of degenerate states and least of all can we speak of its
breakdown. In the light of the analysis of Roy and Sharma\cite{RS19} it
seems that the earlier papers on the HFT\cite{AC03,F04} were not clearly
understood.

The group of unitary operators that carry out the following transformations
of the coordinates $U_{0}:(x,y)\rightarrow (x,y)$ (identity), $%
U_{1}:(x,y)\rightarrow (-x,-y)$, $U_{2}:(x,y)\rightarrow (y,x)$, $%
U_{3}:(x,y)\rightarrow (-y,-x)$ is isomorphic to the well known group $%
C_{2v} $\cite{C90}. They leave the Hamiltonian operator (\ref{eq:HO})
invariant ($U_{i}^{\dagger }HU_{i}=H$). The eigenfunctions (\ref
{eq:HO_psi_mn(0)}) are basis for the irreps of the group $C_{2v}$ but the
eigenfunctions (\ref{eq:HO_varphi}) are not. For example, $U_{2}\phi _{\nu
-i}(\omega ,x)\phi _{i}(\omega ,y)=\phi _{i}(\omega ,x)\phi _{\nu -i}(\omega
,y)$, while $U_{2}\phi _{m}\left( \omega ,q_{1}\right) \phi _{n}\left(
\omega ,q_{2}\right) =(-1)^{n}\phi _{m}\left( \omega ,q_{1}\right) \phi
_{n}\left( \omega ,q_{2}\right) $. The suitable linear combinations of the
eigenfunctions (\ref{eq:HO_varphi}) should be $\frac{1}{\sqrt{2}}\left[
\varphi _{\nu i}(x,y)\pm \varphi _{\nu i}(y,x)\right] $ for $\nu -i\neq i$.

The second example is given by the Hamiltonian matrix
\begin{equation}
\mathbf{H}(\lambda )=\left(
\begin{array}{llllll}
0 & 1 & 0 & 0 & 0 & \lambda \\
1 & 0 & 1 & 0 & 0 & 0 \\
0 & 1 & 0 & \lambda & 0 & 0 \\
0 & 0 & \lambda & 0 & 1 & 0 \\
0 & 0 & 0 & 1 & 0 & 1 \\
\lambda & 0 & 0 & 0 & 1 & 0
\end{array}
\right) ,\;\lambda >0,  \label{eq:mat_H}
\end{equation}
with eigenvalues
\begin{eqnarray}
\epsilon _{1} &=&-\frac{\lambda +\sqrt{\lambda ^{2}+8}}{2},\;\epsilon _{2}=%
\frac{\lambda -\sqrt{\lambda ^{2}+8}}{2},\;\epsilon _{3}=-\lambda ,
\nonumber \\
\epsilon _{4} &=&\lambda ,\;\epsilon _{5}=\frac{\sqrt{\lambda ^{2}+8}%
-\lambda }{2},\;\epsilon _{6}=\frac{\lambda +\sqrt{\lambda ^{2}+8}}{2}.
\label{eq:mat_energies}
\end{eqnarray}
We appreciate that $\epsilon _{2}(1)=\epsilon _{3}(1)$ and $\epsilon
_{4}(1)=\epsilon _{5}(1)$ and in what follows we analyze just the former
crossing. The corresponding eigenvectors are
\begin{equation}
\mathbf{v}_{2}=\frac{\mathbf{w}_{2}}{\sqrt{\mathbf{w}_{2}\cdot \mathbf{w}_{2}%
}},\;\mathbf{w}_{2}=\left(
\begin{array}{c}
1 \\
-\frac{\sqrt{\lambda ^{2}+8}+\lambda }{2} \\
1 \\
1 \\
\frac{\sqrt{\lambda ^{2}+8}-3\lambda }{\lambda \sqrt{\lambda ^{2}+8}-\lambda
^{2}-2} \\
1
\end{array}
\right) ,\;\mathbf{v}_{3}=\frac{1}{2}\left(
\begin{array}{c}
1 \\
0 \\
-1 \\
1 \\
0 \\
-1
\end{array}
\right) .
\end{equation}
One can easily verify that
\begin{equation}
\mathbf{v}_{2}^{t}\cdot \frac{d\mathbf{H}}{d\lambda }\cdot \mathbf{v}_{2}=%
\frac{d\epsilon _{2}}{d\lambda },\;\mathbf{v}_{3}^{t}\cdot \frac{d\mathbf{H}%
}{d\lambda }\cdot \mathbf{v}_{3}=\frac{d\epsilon _{3}}{d\lambda },
\end{equation}
where the superscript $t$ stands for transpose. These expressions are valid
for all $\lambda $ including the crossing point $\lambda =1$. This is
another simple example that confirms our general conclusion given in Sec.~%
\ref{sec:HFT} about the continuity of all the mathematical expressions
related to the HFT.

The Hamiltonian operator (\ref{eq:mat_H}) is invariant under the similarity
transformations given by the following orthogonal matrices
\begin{eqnarray}
\mathbf{E} &=&\left(
\begin{array}{cccccc}
1 & 0 & 0 & 0 & 0 & 0 \\
0 & 1 & 0 & 0 & 0 & 0 \\
0 & 0 & 1 & 0 & 0 & 0 \\
0 & 0 & 0 & 1 & 0 & 0 \\
0 & 0 & 0 & 0 & 1 & 0 \\
0 & 0 & 0 & 0 & 0 & 1
\end{array}
\right) ,\;\mathbf{C}_{2v}=\left(
\begin{array}{cccccc}
0 & 0 & 0 & 1 & 0 & 0 \\
0 & 0 & 0 & 0 & 1 & 0 \\
0 & 0 & 0 & 0 & 0 & 1 \\
1 & 0 & 0 & 0 & 0 & 0 \\
0 & 1 & 0 & 0 & 0 & 0 \\
0 & 0 & 1 & 0 & 0 & 0
\end{array}
\right) ,\;  \nonumber \\
\mathbf{\sigma }_{v1} &=&\left(
\begin{array}{cccccc}
0 & 0 & 0 & 0 & 0 & 1 \\
0 & 0 & 0 & 0 & 1 & 0 \\
0 & 0 & 0 & 1 & 0 & 0 \\
0 & 0 & 1 & 0 & 0 & 0 \\
0 & 1 & 0 & 0 & 0 & 0 \\
1 & 0 & 0 & 0 & 0 & 0
\end{array}
\right) ,\;\mathbf{\sigma }_{v2}=\left(
\begin{array}{cccccc}
0 & 0 & 1 & 0 & 0 & 0 \\
0 & 1 & 0 & 0 & 0 & 0 \\
1 & 0 & 0 & 0 & 0 & 0 \\
0 & 0 & 0 & 0 & 0 & 1 \\
0 & 0 & 0 & 0 & 1 & 0 \\
0 & 0 & 0 & 1 & 0 & 0
\end{array}
\right) .  \label{eq:C_2v_matrices}
\end{eqnarray}
They are realizations of the elements of the group $C_{2v}$\cite{C90}. The
eigenvectors $\mathbf{v}_{1}$, $\mathbf{v}_{2}$, $\mathbf{v}_{3}$, $\mathbf{v%
}_{4}$, $\mathbf{v}_{5}$ and $\mathbf{v}_{6}$ of $\mathbf{H}(\lambda )$ are
basis for the irreps $B_{2}$, $A_{1}$, $A_{2}$, $B_{1}$, $B_{2}$ and $A_{1}$%
, respectively, for $\lambda \neq 1$. In the limit $\lambda \rightarrow 1$
they retain their symmetry and satisfy the diagonal HFT. If one decides to
diagonalize the Hamiltonian $\mathbf{H}(1)$ and use its eigenvectors in an
application of the diagonal HFT one should choose those linear combinations
of the degenerate eigenvectors that are basis for the irreps of $C_{2v}$.
Notice, for example, that any arbitrary linear combination of the
eigenvectors $\mathbf{v}_{2}(\lambda =1)$ and $\mathbf{v}_{3}(\lambda =1)$
chosen above as illustrative examples will not have the correct symmetry
because those eigenvectors are basis for different irreps ($A_{1}$ and $%
A_{2} $, respectively).

\begin{figure}[tbp]
\begin{center}
\includegraphics[width=9cm]{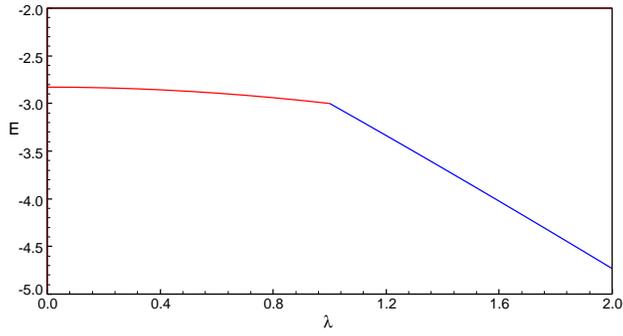}
\end{center}
\caption{Energy given by equation (\ref{eq:fermions_E_0(lambda)})}
\label{Fig:E_0}
\end{figure}

\begin{figure}[tbp]
\begin{center}
\includegraphics[width=9cm]{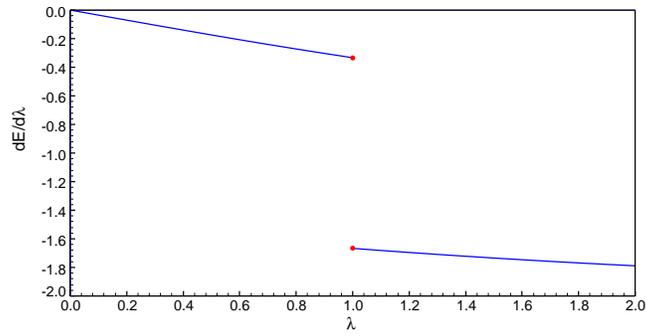}
\end{center}
\caption{ Slope of the energy (\ref{eq:fermions_E_0(lambda)}))}
\label{Fig:dE_0}
\end{figure}

In order to make the discussion of this model closer to the problem
considered by Roy and Sharma\cite{RS19} we assume that the Hamiltonian
matrix (\ref{eq:mat_H}) is a one-particle operator for a system of $N$
fermions. Obviously, we can accommodate a maximum of $N=6$ fermions in this
model and for concreteness we will show results for $N_{p}=2$. The lowest
energy levels are
\begin{equation}
E_{0}^{B_{2}}=\epsilon _{1}+\epsilon _{2},\;E_{0}^{B_{1}}=\epsilon
_{1}+\epsilon _{3}.  \label{eq:fermions_E_0(A1),E_0(A2)}
\end{equation}
Following those authors, the energy of the ground state is
\begin{equation}
E_{0}(\lambda )=\left\{
\begin{array}{c}
E_{0}^{B_{2}},\;\lambda <1 \\
E_{0}^{B_{1}},\;\lambda >1
\end{array}
\right. .  \label{eq:fermions_E_0(lambda)}
\end{equation}
Therefore, we have constructed a function $E_{0}(\lambda )$ with a
discontinuous first derivative. The HFT applies to each of the two-fermion
states $\psi _{0}^{B_{2}}$ and $\psi _{0}^{B_{1}}$%
\begin{eqnarray}
\frac{dE_{0}^{B_{2}}}{d\lambda } &=&\left\langle \psi _{0}^{B_{2}}\right|
\frac{dH}{d\lambda }\left| \psi _{0}^{B_{2}}\right\rangle ,  \nonumber \\
\frac{dE_{0}^{B_{1}}}{d\lambda } &=&\left\langle \psi _{0}^{B_{1}}\right|
\frac{dH}{d\lambda }\left| \psi _{0}^{B_{1}}\right\rangle ,
\label{eq:fermions_dE_0^A}
\end{eqnarray}
and both derivatives are continuous. On the other hand, $dE_{0}/d\lambda $
is discontinuous at $\lambda =1$ but can also be obtained by means of the
HFT if one calculates the derivative at the cusp properly. In fact, at
exactly the cusp we obtain two values of the slope that are the two
eigenvalues of the $2\times 2$ matrix representation of $dH/d\lambda $.
Figure~\ref{Fig:E_0} shows the piecewise-defined energy (\ref
{eq:fermions_E_0(lambda)}) and Figure~\ref{Fig:dE_0} its derivative. The two
red circles in the latter mark the two eigenvalues just mentioned. The
triangles in figure 13 of Roy and Sharma\cite{RS19} are the result of a
calculation that yields wrong slopes as the average of the two true slopes
at each crossing as shown in equation (\ref{eq:H'_ii_varphi}).

\section{Conclusions}

\label{sec:Conclusions}

Throughout this paper we have tried to make it clear that the diagonal HFT (%
\ref{eq:HFT_diag}) is valid in the case of degenerate states as argued in
several papers\cite{AC03,F04,V04,BHM04}. Any discrepancy between the two
ways of calculating the slopes of the energy levels arises from the wrong
choice of the eigenfunctions used in the calculation of the expectation
values at a level crossing. The correct degenerate eigenfunctions are those
that satisfy equation (\ref{eq:H'_ij}). This condition is not a correction
of the HFT as misinterpreted by Roy and Sharma\cite{RS19} because the
functions that satisfy it are given naturally by the continuity equation (%
\ref{eq:continuity}). There is no discontinuity whatsoever at a level
crossing as clearly follows from the continuity equation just mentioned.
However, the definition of the ground-state energy as in equation (\ref
{eq:RS_E_0}) forces a discontinuity in $dE_{0}/d\lambda $ because the energy
$E_{0}$ is given by one state on one side of a level crossing and a
different state on the other side of it. At the cusp generated by the level
crossing one has to choose the correct eigenfunctions that are given by
equation (\ref{eq:H'_ij}). The eigenvalues of the $2\times 2$ (for
simplicity we assume $g_{n}=2$) matrix representation of $dH/d\lambda $ will
give two values of the slope at the cusp which are the result of the
continuity equation for $\lambda \rightarrow \lambda _{0}^{-}$ and $\lambda
\rightarrow \lambda _{0}^{+}$.

The states that cross at some value of the model parameter have different
symmetries\cite{F14} (and references therein). These states obviously
satisfy equation (\ref{eq:H'_ij}). Consequently, if one carries out
calculations for each symmetry species separately no crossing occurs and one
is not forced to construct the degenerate eigenfunctions that satisfy the
HFT (the theorem is automatically satisfied for each irrep). Any arbitrary
linear combination of degenerate eigenfunctions mixes different symmetries
and one obtains the wrong result shown in equation (\ref{eq:H'_ii_varphi}).
An example is given by the triangles in figure 13 of Roy and Sharma\cite
{RS19}. These conclusions are also valid for approximate variational
wavefunctions. Of course this analysis should be carefully applied to the
case in which one is forced (for physical reasons) to choose always the
lowest energy level $E_{0}$ because it is related to one irrep when $\lambda
<\lambda _{0}$ and another one for $\lambda >\lambda _{0}$.

We expect that the present paper makes the issue of the HFT for degenerate
states clearer than the previous one\cite{F04}.

\begin{thebibliography}{99}
\bibitem{F39}  R. P. Feynman, Phys. Rev. \textbf{56}, 340 (1939).

\bibitem{CDL77}  C. Cohen-Tannoudji, B. Diu, and F. Lalo\"{e}, Quantum
Mechanics (John Wiley \& Sons, New York, 1977).

\bibitem{P68}  F. L. Pilar, Elementary Quantum Chemistry (McGraw-Hill, New
York, 1968).

\bibitem{E54}  S. T. Epstein, Am. J. Phys. \textbf{22}, 613 (1954).

\bibitem{BS89}  S. Brajamani Singh and C. A. Singh, Am. J. Phys. \textbf{57}%
, 894 (1989).

\bibitem{ZG02}  G. P. Zhang and T. F. George, Phys. Rev. B \textbf{66},
033110 (2002).

\bibitem{AC03}  O. E. Alon and Cederbaum. L. S., Phys. Rev. B \textbf{68},
033105 (2003).

\bibitem{F04}  F. M. Fern\'{a}ndez, Phys. Rev. B \textbf{69}, 037101 (2004).

\bibitem{V04}  S. R. Vatsya, Phys. Rev. B \textbf{69}, 037102 (2004).

\bibitem{BHM04}  R. Balawender, A. Holas, and N. H. March, Phys. Rev. B
\textbf{69}, 037103 (2004).

\bibitem{RS19}  N. Roy and A. Sharma, Phys. Rev. B \textbf{100}, 195143
(2019).

\bibitem{C90}  F. A. Cotton, Chemical Applications of Group Theory, Third
ed. (John Wiley \& Sons, New York, 1990).

\bibitem{AF19}  P. Amore and F. M. Fern\'{a}ndez, Comment on: ``Tunneling of
coupled methyl quantum rotors in 4-methylpyridine: Single rotor potential
versus coupling interaction''. J. Chem. Phys. \textbf{147}, 194303 (2017),
arXiv:1911.04909 [physics.chem-ph].

\bibitem{F14}  F. M. Fern\'{a}ndez, J. Math. Chem. \textbf{52}, 2322 (2014).
\end{thebibliography}
\end{document}